\documentclass{ws-ijmpe}
\usepackage{graphicx}
\usepackage{epsfig}

\def\case#1/#2{\textstyle\frac{#1}{#2}}

\newcommand{\be}{\begin{equation}}
\newcommand{\ee}{\end{equation}}
\newcommand{\ben}{\begin{eqnarray}}
\newcommand{\een}{\end{eqnarray}}

\def\ex{e_1{}^1}
\def\ey{e_2{}^2}

\def\y{\vartheta}
\def\z{\varphi}

\begin{document}

\markboth{Genly Leon}{Phase space of anisotropic $R^n$
cosmologies}

\catchline{}{}{}{}{}

\title{PHASE SPACE OF ANISOTROPIC $R^n$ COSMOLOGIES 
}

\author{\footnotesize GENLY LEON
}

\address{Department of Mathematics, Universidad Central de Las Villas, Carretera a Camajuani km 5.5\\
Santa Clara, Villa Clara ZIP/Zone 54830, Cuba 
genly@uclv.edu.cu}

\maketitle

\begin{history}
\received{(received date)}
\revised{(revised date)}
\end{history}

\begin{abstract}
We construct general anisotropic cosmological scenarios governed
by an $f(R)=R^n$ gravitational sector. Focusing then on some
specific geometries, and modelling the matter content as a perfect
fluid, we perform a phase-space analysis. We analyze the
possibility of accelerating expansion at late times, and
additionally, we determine conditions for the parameter $n$ for
the existence of phantom behavior, contracting solutions as well
as of cyclic cosmology. Furthermore, we analyze if the universe
evolves towards the future isotropization without relying on a
cosmic no-hair theorem. Our results indicate that anisotropic
geometries in modified gravitational frameworks present radically
different cosmological behaviors compared to the simple isotropic
scenarios.
\end{abstract}

\section{Introduction}

There exist a huge observational evidence that the expansion rate
of our universe is now accelerating \cite{obs}. One way to explain
this feature is to consider the extended gravitational theories
known as $f(R)$-gravity (see \cite{Sotiriou:2008rp} and references
therein). In such approach the Hilbert-Einstein action is
generalized by replacing the Ricci scalar $R$ by functions of it.

According to the observational evidence, the universe, highly
inhomogeneous and anisotropic in earlier epochs, had evolved to
the homogeneous and isotropic state we observe today with great
accuracy. The robust approach to answer this question is to begin
with an initially anisotropic universe and analyze if the universe
evolves towards the future isotropization.

Anisotropic but homogeneous cosmologies was known since a long
time ago \cite{Misner:1974qy}. The most well-studied homogeneous
but anisotropic geometries are the Bianchi type (see
\cite{Ellis:1968vb} and references therein) and the
Kantowski-Sachs metrics \cite{KS}, either in conventional or in
higher-dimensional framework. For Bianchi I, Bianchi III, and
Kantowski-Sachs geometries one can obtain a very good picture of
homogeneous but anisotropic cosmology by using both numerical and
analytical approaches and incorporating also the matter content
(see \cite{Leon:2010pu} and references therein).

In the present work we are interested in investigating the phase
space of anisotropic $f(R)=R^n$ cosmologies focussing in
Kantowski-Sachs geometries and modelling the matter content as a
perfect fluid. The major emphasis is on the late-time stable
solutions. We analyze the possibility of accelerating expansion at
late times, and additionally, we determine conditions for the
parameter $n$ for the existence of phantom behavior, contracting
solutions as well as of cyclic cosmology. Furthermore, we analyze
if the universe evolves towards the future isotropization. We make
few comments about the feasibility to construct compact phase
spaces for Bianchi III and Bianchi I background.

We stress that the results of anisotropic $f(R)$ cosmology are
expected to be different than the corresponding ones of
$f(R)$-gravity in isotropic geometries, similarly to the
differences between isotropic \cite{eddington} and anisotropic
\cite{harrison} considerations in General Relativity.
Additionally, the results are expected to be different from
anisotropic General Relativity, too. As we see, anisotropic $f(R)$
cosmology can be consistent with observations.

\section{Basic Framework}

In order to investigate anisotropic cosmologies, it is usual to
assume an anisotropic metric of the form \cite{Byland:1998gx}:
{\small{
\begin{equation}
 ds^2 = - N(t)^2 dt^2 + [\ex(t)]^{-2} dr^2
  + [\ey(t)]^{-2} [d\y^2 + S(\y)^2\,
   d\z^2],\label{metric}
\end{equation}}}
where $1/\ex(t)$ and $1/\ey(t)$ are the expansion scale factors.

The  metric \eqref{metric} can describe three geometric families,
that is
\begin{eqnarray}
S(\y ) &=&\left\{
\begin{array}{l}
\sin {\y }\hspace{0.7cm}{\rm for}\hspace{0.3cm}k=+1, \\
\y \hspace{1.2cm}{\rm for}\hspace{0.3cm}k=0, \\
\sinh {\y }\hspace{0.5cm}{\rm for}\hspace{0.3cm}k=-1,
\end{array}
\right. \
\end{eqnarray}
known respectively as Kantowski-Sachs, Bianchi I and Bianchi III
models. In the metric formalism for $f(R)$-gravity the
fourth-order cosmological equations write
\cite{Sotiriou:2008rp,goswami,goheer}:
\begin{equation}
 G_{\mu \nu}=\frac{T_{\mu \nu}^{(m)}}{f'(R)}+\frac{1}{f'(R)}\left[\frac{1}{2} g_{\mu
\nu}\left(f(R)-R f'(R)\right)+ \nabla_\mu \nabla_\nu f'(R)-g_{\mu
\nu}\Box f'(R)\right],\label{EFE}
 \end{equation}
 where the prime
denotes differentiation with respect to the Ricci scalar $R$,
$\nabla_\mu$ is the covariant derivative associated to the
Levi-Civita connection of the metric and $\Box\equiv
\nabla^\mu\nabla_\mu.$ $T_{\mu \nu}^{(m)}$ denotes the matter
energy-momentum tensor, which is assumed to correspond to a
perfect fluid with energy density $\rho_m$ and pressure $p_m$. The
Ricci scalar is given by $R= 12H ^{2}+6 \sigma^{2}+6\dot H +2K k,$
where $k=+1, 0, -1$ corresponds to Kantowski-Sachs, Bianchi I and
Bianchi III models respectively. In this equation we have
introduced the kinematical variable $\sigma=\frac13\frac{  d}{ {d}
t}\left[\ln
      \frac{\ex}{\ey}\right]$ and the Hubble scalar $ H=-\frac13\frac{  d}{ {d} t}\left[\ln \ex
(\ey)^2\right].$ $K \equiv (\ey)^2$ is the Gauss curvature of the
3-spheres (see \cite{carge73} for the general formalism to deduce
kinematical variables  and \cite{Leon:2010pu} for an explicit
computation in such geometrical backgrounds).

Now we impose an ansatz of the form $f(R)=R^n$
\cite{Sotiriou:2008rp,Carloni05,Goheer:2007wu}, since such an
ansatz does not alter the characteristic length scale (and General
Relativity is recovered when $n=1$), and it leads to simple exact
solutions which allow for comparison with observations
\cite{Capozziello}. Additionally, following \cite{goswami,goheer}
we consider that the parameter $n$ is related to the matter
equation-of-state parameter through $ n=\frac{3}{2}(1+w),$ with
$w=\frac{p_m}{\rho_m},$ where $\rho_m$ and $p_m$ are the energy
density and pressure of the matter perfect fluid. Imposing all the
energy conditions for standard matter follows $-1/3\leq w\leq1$
($n\in[1,3]$). The most interesting cases being those of
 dust ($w=0$, $n=3/2$) and radiation fluid
($w=1/3$, $n=2$).

The cosmological equations of $R^n$-gravity  in the
Kantowski-Sachs, Bianchi I and Bianchi III backgounds are: the
``Raychaudhuri equation'' \eqref{cosmeq1}, the shear evolution
\eqref{cosmeq2}, the trace equation \eqref{cosmeq4} (obtained from
equation (5) in section IIA of \cite{Capozziello:2009nq}), the
Gauss constraint \eqref{Gauss2}, the evolution equation for the
2-curvature $K$ \eqref{propK}, as well as the matter conservation
equation \eqref{cosmeq5} :

\begin{eqnarray}
 &\label{cosmeq1}
  \dot H+ H^2=
-2\sigma^2-\frac{\rho_m}{3 n R^{n-1}}+\frac{1}{6n} R+(n-1)
H\frac{\dot R}{R},\\
&\label{cosmeq2} \dot\sigma=-\sigma^2-3 H \sigma +H^2
-\frac{\rho_m}{3 n R^{n-1}}-\frac{n-1}{6 n} R
-(n-1)\left(\sigma-H\right)\frac{\dot R}{R},
\\
&\label{cosmeq4} \frac{\ddot R}{R}=\frac{n-2}{3 n(n-1)}
R+\frac{2(2-n)}{3 n(n-1)}\frac{\rho_m}{R^{n-1}}-3 H\frac{\dot
R}{R}-(n-2)\frac{\dot R^2}{R^2}\\
&\label{Gauss2} \left(H+\frac{n-1}{2}\frac{\dot
R}{R}\right)^2+\frac{k}{3} K=\sigma^2+\frac{\rho_m}{3 n R^{n-1}}+
\frac{n-1}{6n} R+\frac{(n-1)^2}{4}\frac{\dot R^2}{R^2},
\\
&\label{propK} \dot {K}=-2 (\sigma +H) \left( {K}\right).
 \\
& \label{cosmeq5}
 \dot \rho_m=-2n H \rho_m,
\end{eqnarray}
In the former we use the notation $k=+1, 0, -1$ for
Kantowski-Sachs, Bianchi I and Bianchi III models respectively.
For Kantowski-Sachs geometry the field equations are the same  as
in (35)-(38), (21), (22) and (39) in \cite{Leon:2010pu}. For
Bianchi I, equation \eqref{propK} decouples.

\section{Phase space analysis}

The so-called Hubble-normalized variables together with a
Hubble-normalized time variable have been used successfully to
study the isotropization of cosmological models \cite{Collins}.
For simple classes of ever expanding models such as the open and
flat FLRW models and the spatially homogeneous Bianchi type I
models in GR the dynamical systems variables are bounded even
close to the cosmological singularity \cite{Wainwright04}. These
simple classes of cosmological models do not allow for bouncing,
recollapsing or static models, since there are no contributions to
the Friedmann equation that would allow for the Hubble-parameter
to vanish. In models allowing $H$ to pass through zero (e.g. the
simple addition of positive spatial curvature), the state space
obtained from expansion normalized variables becomes non-compact
(see \cite{Leach:2006br} for such an analysis in Bianchi I
$R^n$-gravity). Hence, one has to perform an additional analysis
to study the equilibrium points at infinity using, for instance,
the well-known Poincar\'e projection \cite{Poincare}, where the
points at infinity are projected onto a unit sphere.
Alternatively, one may break up the state space into compact
subsectors, where the dynamical systems and time variables are
normalized differently in each sector (see e.g. \cite{Goliath99}
and \cite{Goheer:2007wx} for Bianchi I; see also
\cite{Goheer:2007wu} for a comprehensive analysis of Bianchi III,
Bianchi I and Kantowski-Sachs geometries in $R^n$-gravity). The
full state space is then obtained by pasting the compact
subsectors together. Both methods have advantages and
disadvantages in this context \cite{Goheer:2007wx}.

\subsection{Kantowski-Sachs}

The phase space of Kantowski-Sachs geometry in $R^n$ gravity was
investigated in \cite{Leon:2010pu} by using the auxiliary
variables \cite{Goheer:2007wx,Solomons:2001ef}: $Q=\frac{H}{D},
\Sigma=\frac{\sigma}{D}, x=\frac{(n-1) \dot R}{2 R D},
y=\frac{(n-1)R}{6 n D^2},
 z=\frac{\rho_m}{3 n R^{n-1}D^2}, {\cal K}=\frac{K}{3 D^2},
$ where $D=\sqrt{\left(H+\frac{n-1}{2}\frac{\dot
R}{R}\right)^2+\frac{1}{3} K},$ and the time variable $\tau$
through ${d}\tau=\left(\frac{D}{n-1}\right) {d} t.$

The restrictions $
 x^2+y+z+\Sigma^2=1,$ and $ \left(Q+x\right)^2+K=1,$ enable
us to investigate the reduced dynamical system in the variables
$\left(Q,\Sigma,x,y\right).$

In tables \ref{tab1} and \ref{tab2} we present partial information
about the isolated and curves of critical point of corresponding
vector field.
\begin{table*}
    \centering
    \caption{\label{tab1}The isolated critical points of the
cosmological system.
  We use the notations  $n_+=\frac{1+\sqrt{3}}{2}\approx 1.37$,
  $Q^\star=\frac{2n^2-5n+5}{7n^2-16n+10},$ $\Sigma^\star=-\frac{2n^2-2n-1}{7n^2-16n+10},$
   $x^\star=\frac{3(n-1)(n-2)}{7n^2-16n+10},$ and $y^\star=\frac{9(4n^2-10n+7)(n-1)^2}{(7n^2-16n+10)^2}$.}
\begin{tabular}{cccccc}
  \hline   \hline
  Name & $Q$ & $\Sigma$ & $x$ & $y$ & Existence\\
  \hline \hline
  $\mathcal{A}_+$ & $\frac{2n-1}{3(n-1)}$ & $0$ & $\frac{n-2}{3(n-1)}$
  & $\frac{(2n-1)(4n-5)}{9(n-1)^2}$ & $\frac{5}{4} \leq n \leq 3$
 \\ \hline
  $\mathcal{A}_-$ & ${-}\frac{2n-1}{3(n-1)}$ & $0$ & $-\frac{n-2}{3(n-1)}$
  & $\frac{(2n-1)(4n-5)}{9(n-1)^2}$ & $\frac{5}{4} \leq n \leq 3$
 \\ \hline
  ${\cal B}_+$ & $\frac{1}{3-n}$ & $0$ & $\frac{n-2}{n-3}$  & $0$ & $1< n\leq \frac{5}{2}$\\ \hline
    ${\cal B}_-$ & $-\frac{1}{3-n}$ & $0$ & $-\frac{n-2}{n-3}$  & $0$ & $1< n\leq \frac{5}{2}$\\ \hline
  $P_3^+$ & $\frac{1}{2-n}$ & $0$ & $-\frac{n-1}{2-n}$ & $\frac{n-1}{n(n-2)^2}$
  & $1< n\leq n_+$\\ \hline
  $P_4^+$ & $Q^\star$ & $\Sigma^\star$
  & $x^\star$  & $y^\star$ & $n_+\leq n\leq 3$
  \\ \hline
  $P_4^-$ & $-Q^\star$ & $-\Sigma^\star$
  & $-x^\star$  & $y^\star$ & $n_+\leq n\leq 3$
  \\ \hline
  $P_5^+$ & $\frac{1}{n-2}$
  & $\frac{\sqrt{2n-5}}{n-2}$ & $\frac{n-3}{n-2}$ & $0$ & $\frac{5}{2}\leq n\leq 3$ \\ \hline
  $P_5^-$ & $-\frac{1}{n-2}$
  & $\frac{\sqrt{2n-5}}{n-2}$ & $-\frac{n-3}{n-2}$ & $0$ & $\frac{5}{2}\leq n\leq 3$  \\ \hline
  \hline
\end{tabular}
\end{table*}

\begin{table*}
    \centering
    \caption{\label{tab2}The curves of critical points $C_\epsilon$,
and the representative critical points ${\cal N}_\epsilon,$
$\mathcal{L}_\epsilon$, $P_1^\epsilon$ and $P_2^\epsilon$ of the
cosmological system (we use $\epsilon=\pm1$). $u$ varies in
$[0,2\pi].$}
    \begin{tabular}{ccccccc}
  \hline   \hline
  Name & $Q$ & $\Sigma$ & $x$ & $y$ & Existence& Stability\\
  \hline \hline
  $\mathcal{N_+}$ &  0  & 0  & 1
  & 0 & always & unstable
 \\ \hline
  $\mathcal{N_-}$ &  0  &  0 & -1
  & 0 & always & stable
 \\ \hline
    $\mathcal{L_+}$ &  2  &  0 & -1
  & 0 & always & unstable
 \\ \quad & \quad & \quad & \quad & \quad & \quad & for $\frac{5}{4}\leq n\leq\frac{5}{2}$\\\hline
  $\mathcal{L_-}$ &  -2  &  0 & 1
  & 0 & always & stable
 \\  \quad & \quad & \quad & \quad & \quad & \quad & for $\frac{5}{4}\leq n\leq\frac{5}{2}$\\\hline
      $P_1^+$ & $1$ & $1$ & $0$ & $0$ & $1< n\leq 3$ &{\text{non-hyperbolic }}\\ 
      \quad & \quad & \quad & \quad & \quad & \quad &{\text{$3D$ unstable manifold}}\\ \hline
  $P_1^-$ & $-1$ & $-1$ & $0$ & $0$ & $1< n\leq 3$ &{\text{non-hyperbolic \ \ \ }}\\ 
  \quad & \quad & \quad & \quad & \quad & \quad &{\text{$3D$ stable manifold}}\\ \hline
  $P_2^+$ & $1$ & $-1$ & $0$ & $0$ & $1< n\leq 3$ &{\text{non-hyperbolic}} \\ 
  \quad & \quad & \quad & \quad & \quad & \quad &{\text{$3D$ unstable manifold}}\\ \hline
   $P_2^-$ & $-1$ & $1$ & $0$ & $0$ & $1< n\leq 3$ &{\text{non-hyperbolic \ \ \  }}\\ 
   \quad & \quad & \quad & \quad & \quad & \quad &{\text{$3D$ stable manifold}}\\ \hline
  ${\cal C}_+$ & $1+\sin u$ &  $\cos u$ & $-\sin u$ & 0  & always & unstable if   \\
  \quad & \quad & \quad & \quad & \quad & \quad & $\frac{5}{4}\leq n\leq\frac{5}{2}$\\\hline
    ${\cal C}_-$ & $-1+\sin u$ &  $\cos u$ & $-\sin u$ & 0  & always & stable if \\
    \quad & \quad & \quad & \quad & \quad & \quad & $\frac{5}{4}\leq n\leq\frac{5}{2}$\\\hline
  \hline
\end{tabular}
\end{table*}

The curves $C_\epsilon$ contain the representative critical points
 ${\cal N}_\epsilon$ and
$\mathcal{L}_\epsilon$ described in \cite{goheer} (see table
\ref{tab2}). Additionally to the investigation of \cite{goheer},
we obtain four new critical points, which are a pure result of the
anisotropy: $P_{1,2}^\epsilon.$ For $-\frac{1}{6}<w<\frac{2}{3},$
${\cal L}_+$ is unstable and ${\cal L_-}$ is stable.
 The critical point ${\cal
N}_+$ (${\cal N}_-$) is always unstable (stable) except in the
case $w= -\frac{1}{3}.$ These results match with the results
obtained in \cite{goheer}. In this scenario there are also
contracting solutions, either accelerating or decelerating, which
are, in general, not globally but locally asymptotically stable.
For  $2<n<3$, the critical points $P_{2,1}^-$ correspond to
accelerating contraction, in which the total matter/energy behaves
like radiation. They do possess a $3D$ stable manifold and their
center manifolds are locally asymptotically stable \cite{wiggins},
and thus they can attract the universe at late times (with small
probability) for this $n$-range.

The stability of the isolated critical points is as follows:
$\mathcal{A}_+$ is stable for $n_+< n < 3$ and a saddle otherwise;
${\cal B}_+$ is a saddle whenever exist; saddle also is $P_4^+$
with a 3-dimensional stable manifold; $P_3^+$ is non-hyperbolic
with a 3-dimensional stable manifold; and $P_5^+$ is
non-hyperbolic, with a 2-dimensional center manifold. The
stability of the critical points in the negative ($-$) branch is
the time reverse of the corresponding points in the positive ($+$)
branch (see further details in table I of \cite{Leon:2010pu}).

As discussed in \cite{Leon:2010pu}, for $1.37\lesssim n<3,$ the
universe at late times can result to a state of accelerating
expansion represented in the phase space by the critical point
$\mathcal{A}_+.$ In the above critical point the isotropization
has been achieved. For $2<n<3$ it exhibits phantom behavior.
Moreover, in the case of radiation ($n=2$, $w=1/3$) the
aforementioned stable solution corresponds to a de-Sitter
expansion describing the inflationary epoch of the universe. These
results were achieved without resort to the cosmic no-hair theorem
\cite{Wald}.  This is not a new feature since de-Sitter solutions
are known to exist in Bianchi I and Bianchi III $R^n$-gravity
\cite{Goheer:2007wu,Leach:2006br}. The acquisition of such a
solution is of great interest, as of many works on anisotropic
cosmologies, since, it is the only robust approach in confronting
isotropy of standard cosmology. The fact that this solution is
accompanied by acceleration or phantom behavior, makes it a very
good candidate for the description of the observable universe
\cite{obs}.

\begin{figure}[!]
\begin{center}
\includegraphics[width=8cm, height=8cm,angle=0]{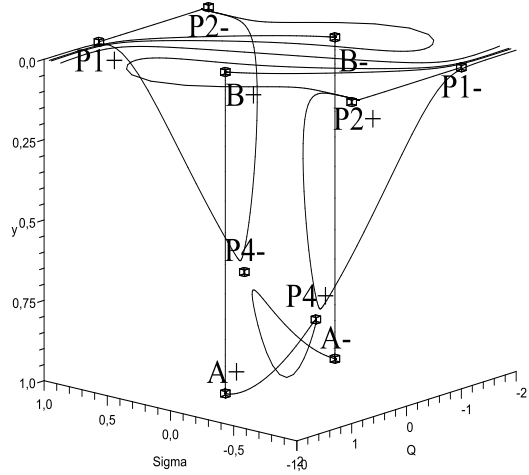}
\label{fig2a}
\caption{Projection of the phase space on the
invariant set $x=0$, in the case of radiation ($w=1/3, n=2$).
There is one orbit of type ${\cal B}_+\longrightarrow {\cal A}_+$
and one of type ${\cal A}_-\longrightarrow {\cal B}_-$. The
existence of an heteroclinic sequence of type ${\cal
A}_-\longrightarrow P_4^-\longrightarrow P_4^+\longrightarrow
{\cal A}_+$ corresponds to a cosmological bounce.}
\end{center}
\end{figure}

In Fig. \ref{fig2a} we display some orbits in the case of
radiation, i.e. for $w=1/3, n=2$, (taken from \cite{Leon:2010pu}). We note the existence of heteroclinic
sequences of types
\begin{eqnarray}
&&{\cal A}_-\longrightarrow \left\{\begin{array}{c}
  P_4^-\longrightarrow P_4^+\longrightarrow {\cal A}_+ \\
  {\cal B}^- \ \ \ \ \ \ \ \ \ \ \ \ \ \ \ \ \ \ \ \ \ \ \ \\
\end{array}\right.\nonumber\\
&&{\cal B}_+\longrightarrow {\cal A}_+\nonumber\\
 &&P_1^-\longrightarrow
P_4^+\longrightarrow P_2^+\nonumber\\
 &&P_2^-\longrightarrow
P_4^-\longrightarrow P_1^+\nonumber\\
 &&P_1^+\longrightarrow P_2^-\nonumber\\
&&P_2^+\longrightarrow P_1^-,\label{heteroclinic}
\end{eqnarray}
revealing the realization of a cosmological bounce or a
cosmological turnaround.

Similarly to the isotropic case (see figure 5 of \cite{goheer})
there is one orbit of type ${\cal B}_+\longrightarrow {\cal A}_+$
and one of type ${\cal A}_-\longrightarrow {\cal B}_-$. However,
in the present case we have the additional existence of an
heteroclinic sequence of type ${\cal A}_-\longrightarrow
P_4^-\longrightarrow P_4^+\longrightarrow {\cal A}_+$,
corresponding to the transition from collapsing AdS to expanding
dS phase, that is we obtain a cosmological bounce followed by a de
Sitter expansion, which could describe the inflationary stage.
This significant behavior is a pure result of the anisotropy and
reveals the capabilities of the scenario.  Lastly, note the very
interesting possibility, of the eternal transition
$P_1^-\longrightarrow P_2^+\longrightarrow P_1^-\longrightarrow
P_2^+\cdots$ which is just the realization of cyclic cosmology
\cite{cyclic}. Bouncing solutions are found to exist both in FRW
$R^n$-gravity \cite{Carloni:2005ii}, as well as in the Bianchi I
and Bianchi III $R^n$-gravity \cite{Goheer:2007wu} (see also
\cite{Barragan:2010qb}), and thus they arise from the $R^n$
gravitational sector.

\section{Conclusions}

As we saw, the universe at late times can result to a state of
accelerating expansion, and additionally, for a particular
$n$-range ($2<n<3$) it exhibits phantom behavior. Additionally,
the universe has been isotropized, independently of the anisotropy
degree of the initial conditions, and it asymptotically becomes
flat. The fact that such features are in agreement with
observations \cite{obs} is a significant advantage of the model.
Moreover, in the case of radiation ($n=2$, $w=1/3$) the
aforementioned stable solution corresponds to a de-Sitter
expansion, and it can also describe the inflationary epoch of the
universe. In our work we extracted our results without relying at
all on the cosmic no-hair theorem, which is a significant
advantage of the analysis \cite{Wald}.

The Kantowski-Sachs anisotropic $R^n$-gravity can also lead to
contracting solutions, either accelerating or decelerating, which
are not globally stable. Thus, the universe can remain near these
states for a long time, before the dynamics remove it towards the
above expanding, accelerating, late-time attractors. The duration
of these transient phases depends on the specific initial
conditions.

One of the most interesting behaviors is the possibility of the
realization of the transition between expanding and contracting
solutions during the evolution. That is, the scenario at hand can
exhibit the cosmological bounce or turnaround. Additionally, there
can also appear an eternal transition between expanding and
contracting phases, that is we can obtain cyclic cosmology. These
features can be of great significance for cosmology, since they
are desirable in order for a model to be free of past or future
singularities.


\begin{thebibliography}{0}

\bibitem{obs}
M.~Kowalski {\it et al.}, Astrophys.\ J.\  {\bf 686}, (2008) 749;
S.~W.~Allen, D.~A.~Rapetti, R.~W.~Schmidt, H.~Ebeling, G.~Morris and A.~C.~Fabian, Mon.\ Not.\ Roy.\ Astron.\ Soc.\  {\bf 383}, (2008) 879;
K.~N.~Abazajian {\it et al.}, Astrophys.\ J.\ Suppl.\  {\bf 182}, (2009) 543;
N.~Jarosik {\it et al.}, Astrophys. J. Suppl. {\bf 192} (2011) 14. 
\bibitem{Sotiriou:2008rp}
T.~P.~Sotiriou and V.~Faraoni, Rev.\ Mod.\ Phys.\  {\bf 82}, (2010) 451;
A.~De Felice and S.~Tsujikawa, Living Rev.\ Rel.\  {\bf 13}, (2010) 3. 
\bibitem{Misner:1974qy}
C.~W.~Misner, K.~S.~Thorne and J.~A.~Wheeler,  {\it{Gravitation}}, San Francisco, W. H. Freeman \& Co. (1973);
P.~J.~E.~Peebles, {\it Principles of physical cosmology}, Princeton, USA: Univ. Pr. (1993) 718 p.;
\bibitem{Ellis:1968vb} 
G.~F.~R.~Ellis and M.~A.~H.~MacCallum, Commun.\ Math.\ Phys.\  {\bf 12}, (1969) 108;
C.~G.~Tsagas, A.~Challinor and R.~Maartens, Phys.\ Rept.\  {\bf 465}, (2008) 61.
\bibitem{KS}
A.S. Kompaneets and A.S. Chernov, Zh. Eksp. Teor. Fiz. {\bf 47}, (1964) 1939. 
English translation: Soviet. Phys. JETP {\bf 20}, (1965) 1303;
R.~Kantowski and R.~K. Sachs, J. Math. Phys. {\bf 7} 443 (1966);
C.~B.~Collins, J.\ Math.\ Phys.\  {\bf 18} (1977) 2116;
E.~Weber, J.\ Math.\ Phys.\  {\bf 25}, (1984) 3279;
O.~Gron, J.\ Math.\ Phys.\  {\bf 27}, (1986) 1490;
M.~Demianski, Z.~A.~Golda, M.~Heller and M.~Szydlowski, Class.\ Quant.\ Grav.\  {\bf 5}, (1988) 733;
L.~Bombelli and R.~J.~Torrence, Class.\ Quant.\ Grav.\  {\bf 7}, (1990) 1747;
L.~M.~Campbell and L.~J.~Garay, Phys.\ Lett.\  B {\bf 254}, (1991) 49;
L.~E.~Mendes and A.~B.~Henriques, Phys.\ Lett.\  B {\bf 254}, (1991) 44;
P.~Vargas Moniz, Phys.\ Rev.\  D {\bf 47}, (1993) 4315;
M.~Cavaglia, Mod.\ Phys.\ Lett.\  A {\bf 9}, (1994) 1897;
S.~Nojiri, O.~Obregon, S.~D.~Odintsov and K.~E.~Osetrin, Phys.\ Rev.\  D {\bf 60}, (1999) 024008;
A.~K.~Sanyal, Phys.\ Lett.\  B {\bf 524}, (2002) 177;
X.~Z.~Li and J.~G.~Hao, Phys.\ Rev.\  D {\bf 68}, (2003) 083512;
W.~F.~Kao, Phys.\ Rev.\  D {\bf 74}, (2006) 043522.
\bibitem{Leon:2010pu}
  G.~Leon, E.~N.~Saridakis, Class.\ Quant.\ Grav.\  {\bf 28}, (2011) 065008.
\bibitem{eddington}
A. S. Eddington, Mon.\ Not.\ Roy.\ Astron.\ Soc.\  {\bf 90}, (1930) 668.
\bibitem{harrison}
E. R. Harrison, {\it Rev. Mod. Phys.} {\bf 39}, (1967) 862;
G. W. Gibbons, {\it Nucl. Phys.} B{\bf 292}, (1987) 784 ; 
J. D. Barrow, G. F. R.  Ellis, R.  Maartens and C. Tsagas {\it Class. Quant. Grav.} {\bf 20}, (2003) L155.
\bibitem{Byland:1998gx}
S.~Byland and D.~Scialom,  Phys.\ Rev.\  D {\bf 57}, (1998) 6065;
U.~Camci, I.~Yavuz, H.~Baysal, I.~Tarhan and I.~Yilmaz, Int.\ J.\ Mod.\ Phys.\  D {\bf 10},(2001) 751;
A.~A.~Coley, W.~C.~Lim and G.~Leon, arXiv:0803.0905 [gr-qc].
\bibitem{goswami}
R.~Goswami, N.~Goheer and P.~K.~S.~Dunsby, Phys.\ Rev.\  D {\bf 78}, (2008) 044011.
\bibitem{goheer}
N.~Goheer, R.~Goswami and P.~K.~S.~Dunsby, Class.\ Quant.\ Grav.\  {\bf 26}, (2009) 105003.
\bibitem{carge73}   G. F. R. Ellis, \textit{Carg\`{e}se Lectures in Physics,}
\textit{Vol} \textbf{6} (ed) E Scatzman  (New York: Gordon and
Breach) (1973);
G. F. R. Ellis   and H. van Elst, \textit{Cosmological
Models (Carg\`{e}se Lectures 1998), Theoretical and Observational
Cosmology} (ed) M. Lachi\`{e}ze-Rey (Kluwer, Dordrecht) (1999) 1-116; 
H.~van Elst and C.~Uggla, Class.\ Quant.\ Grav.\  {\bf 14}, (1997) 2673.
\bibitem{Carloni05}
S.~Carloni, P.~K.~S.~Dunsby, S.~Capozziello and A.~Troisi, Class.\ Quant.\ Grav.\  {\bf 22}, (2005) 4839.
\bibitem{Goheer:2007wu}
N.~Goheer, J.~A.~Leach and P.~K.~S.~Dunsby, Class.\ Quant.\ Grav.\  {\bf 24}, (2007) 5689.
\bibitem{Capozziello}
S.~Capozziello, Int.\ J.\ Mod.\ Phys.\  D {\bf 11}, (2002) 483;
S.~Capozziello, S.~Carloni and A.~Troisi, Recent Res.\ Dev.\ Astron.\ Astrophys.\  {\bf 1}, (2003) 625.
\bibitem{Capozziello:2009nq}
S.~Capozziello, M.~De Laurentis and V.~Faraoni,
arXiv:0909.4672 [gr-qc].
\bibitem{Collins} C. B. Collins {\it Commun. Math. Phys.} {\bf 23} (1971) 137;
C. B. Collins and S. W. Hawking, Astrophys. J., Vol. {\bf 180}, (1973) 317.
\bibitem{Wainwright04} J. Wainwright and W. C. Lim, J. Hyperbol. Diff. Equat. {\bf 2} (2005) 437.
\bibitem{Leach:2006br}
J.~A.~Leach, S.~Carloni and P.~K.~S.~Dunsby, Class.\ Quant.\ Grav.\  {\bf 23}, (2006) 4915.
\bibitem{Poincare} H. Poincar\'{e} {\it J. Math\'{e}matiques}
{\bf 7} (1881) 375;
L. Perko, {\it Differential equations and
dynamical systems}, New York: Springer Verlag (1996).
\bibitem{Goliath99} M. Goliath  and G. F. R. Ellis, Phys. Rev. D. {\bf 60}, (1999)
023502.
\bibitem{Goheer:2007wx}
N.~Goheer, J.~A.~Leach and P.~K.~S.~Dunsby, Class.\ Quant.\ Grav.\  {\bf 25}, (2008) 035013.
\bibitem{Solomons:2001ef}
D.~M.~Solomons, P.~Dunsby and G.~Ellis,  Class.\ Quant.\ Grav.\  {\bf 23}, (2006) 6585.
\bibitem{wiggins} S. Wiggins, {\it{Introduction to Applied Nonlinear Dynamical Systems and Chaos}},
Springer (2003).
\bibitem{Wald}
R.~M.~Wald, Phys.\ Rev.\  D {\bf 28}, (1983) 2118.
\bibitem{cyclic}
P.~J.~Steinhardt and N.~Turok, Phys.\ Rev.\  D {\bf 65}, (2002) 126003;
S.~Mukherji and M.~Peloso, Phys.\ Lett.\  B {\bf 547}, (2002) 297;
J.~Khoury, P.~J.~Steinhardt and N.~Turok, Phys.\ Rev.\ Lett.\  {\bf 92}, (2004) 031302;
L.~Baum and P.~H.~Frampton, Phys.\ Rev.\ Lett.\  {\bf 98}, (2007) 071301;
E.~N.~Saridakis, Nucl.\ Phys.\  B {\bf 808}, 224 (2009);
Y.~F.~Cai and E.~N.~Saridakis, Class.Quant.Grav. 28 (2011) 035010. 
\bibitem{Carloni:2005ii}
S.~Carloni, P.~K.~S.~Dunsby and D.~M.~Solomons, Class.\ Quant.\ Grav.\  {\bf 23}, 1913 (2006).
\bibitem{Barragan:2010qb}
C.~Barragan and G.~J.~Olmo, Phys.\ Rev.\  D {\bf 82}, 084015 (2010).
\end{thebibliography}
\end{document}